# Probing Mid-Infrared Phonon Polaritons in the Aqueous Phase

Haomin Wang,[1] Eli Janzen,[2] Le Wang,[1] James H. Edgar,[2] and Xiaoji G. Xu[1]*

[1]Department of Chemistry, Lehigh University, Bethlehem, PA 18015, USA

[2]Tim Taylor Department of Chemical Engineering, Kansas State University, Durland Hall, Manhattan, KS 66506, USA

*Email: xgx214@lehigh.edu

**Abstract:**

Phonon polaritons (PhPs), the collective phonon oscillations with hybridized electromagnetic fields, concentrate optical fields in the mid-infrared frequency range that matches the vibrational modes of molecules. The utilization of PhPs holds the promise for chemical sensing tools and polariton-enhanced nanospectroscopy. However, investigations and innovations on PhPs in the aqueous phase remains stagnant, because of the lack of *in situ* mid-infrared nano-imaging methods in water. Strong infrared absorption from water prohibits optical delivery and detection in the mid-infrared for scattering-type near-field microscopy. Here, we present our solution: the detection of photothermal responses caused by the excitation of PhPs by liquid phase peak force infrared (LiPFIR) microscopy. Characteristic interference fringes of PhPs in $^{10}$B isotope-enriched *h*-BN were measured in the aqueous phase and their dispersion relationship extracted. LiPFIR enables the measurement of mid-infrared PhPs in the fluid phase, opening possibilities, and facilitating the development of mid-IR phonon polaritonics in water.

**Introduction**

Phonon polaritons (PhPs), the coupling and hybridization of collective lattice vibrations with the local electromagnetic field, lead to high field concentration in the mid-infrared (mid-IR) frequencies. The creation and detection PhPs provide a route for nanoscale infrared sensing of molecular analytes,[1-2] as well as polariton-enhanced nanospectroscopy.[3-4] Chemical sensing of molecules should preferably be performed under the liquid phase—ideally in the water where most of the meaningful chemical reactions and biological transformation happen. Mass transportation in the liquid phase is higher than in air, allowing molecules to reach specific locations where phonon polaritons are active. However, nanoscale *in situ* probing PhPs is currently only performed in air.[5-8] The popular tool for detecting PhPs,[9] scattering scanning near-field optical microscopy (s-SNOM), based on detecting near-field light scattering by atomic force microscopy (AFM), does not straightforwardly operate in water. The lack of a suitable liquid phase compatible nano-imaging technique for PhPs restricts the pace of innovation on mid-IR polaritonics.

The challenge of s-SNOM for probing mid-infrared PhPs in aqueous phase stems from several aspects. First, water strongly absorbs mid-infrared radiation. Thus free-space delivery of mid-IR laser to the tip-sample junction, where near-field light scattering happens, is challenging. Second, s-SNOM requires optical detection of light scattered from the metallic AFM tip, which inevitably passes through water *en route* to the detector, strongly attenuating the signal. Third, the fluid creates a mechanical drag on the AFM cantilever that makes its oscillation anharmonic in the tapping mode feedback.[10] However, the signal extraction of s-SNOM requires a pure harmonic oscillation of the AFM cantilevers. Anharmonicity in the cantilever oscillations causes signal artifacts in the lock-in signal extraction of s-SNOM signals that requires pure harmonic cantilever oscillation. Similarly, if the incident optical field at the tip-sample region is non-uniform, e.g., in an evanescent field from total internal reflection illumination, even purely harmonic oscillation of the AFM cantilever can lead to background signals in the non-fundamental lock-in demodulation. Although s-SNOM can measure water-encapsulated biological samples covered with a layer of graphene in the air,[11] this bypassing approach is associated with complex operational procedures and is highly situational for certain types of samples. The direct and straightforward measurement in the aqueous phase for s-SNOM remains a challenge.

Probing PhPs in the aqueous phase requires a new *in situ* nano-imaging method that overcomes the above challenges and limitations. In this letter, we present our solution to this challenge: the liquid phase peak force infrared (LiPFIR) microscopy, a photothermal actioned-based infrared imaging method that works in the fluid phase. Here, LiPFIR reveals the characteristic frequency-dependent interference fringes from hyperbolic phonon polaritons in [10]B isotope enriched hexagonal boron nitride (*h*-[10]BN) single crystal flakes submerged in water, from which the momentum-energy dispersion relationship of the PhPs is extracted.

It combines the recently developed multipulse peak force infrared (PFIR) microscopy with total internal reflection beam delivery in the liquid phase.[12-13] The scheme of the apparatus of LiPFIR is illustrated in Figure 1a. A liquid-phase AFM (Bioscope Catalyst, Bruker) with a fluid chamber is operated under the peak force tapping mode at the peak force tapping (PFT) frequency of 1 kHz by an AFM controller (Nanoscope V, Bruker). The z-piezo of the AFM probe sinusoidally oscillates above the sample, and the AFM tip intermittently taps into on the sample surface. A lock-in amplifier (MFLi, Zurich Instrument) converts the sinusoidal waveform of the z-piezo drive signal into a phase synchronized square waveform to trigger a function generator (HDG2022B, Hantek). A train of TTL pulses for every PFT cycle is generated with an adjustable time delay by the function generator and routed to trigger a quantum cascade laser (QCL, MIRcat-QT, Daylight Solutions). The infrared laser pulses are guided and focused by a germanium lens (a focal length of 4 cm) into a germanium prism of 20° angle. The 20° angle is slightly larger than the critical angle between the germanium/water interface, so total internal reflection (TIR) is generated at the germanium/water interface with an evanescent field. The metal-coated AFM tip (HQ:NSC14/Cr-Au Mikromasch) further enhances and confines the electromagnetic field of the evanescent field, which excites the polaritonic resonance in the samples. The AFM tip mechanically probes the photothermal expansion of the sample through vertical deflection of the AFM cantilever. The vertical deflection of the cantilever is detected by a quadrant photodiode of a laser beam from a diode laser. The vertical deflection signal is converted into an electrical waveform by a data acquisition card (PXI-5122, National Instruments). Figure 1b displays the cantilever deflection waveform with laser excitation (blue curve). The temporal timing of the laser pulses (red curve) is set when the tip and the sample are in dynamic contact. A fitting procedure is utilized to extract the cantilever oscillations from the slow varying cantilever deflection from the dynamic tapping of the AFM tip. The extracted cantilever oscillation is shown in Figure 1c. Fast Fourier transform recovers the oscillation amplitude that is recorded as the LiFPIR signal. In our experiment, the repetition rate of the pulse train matches one of the cantilever resonant frequencies in the fluid, leading to a moderate amplification of the mechanical detection of photothermal expansion signal. The LiPFIR image is formed by scanning the AFM tip over the sample at a desirable frequency; the LiPFIR spectra are collected by placing the AFM tip at desirable locations and sweep the frequency of the infrared laser source.

The field strength of the evanescent wave of mid-infrared radiation at the germanium/water interface is enhanced by the total internal reflection near the critical angle. Figure 1d displays the finite element simulation of the field strength at the end of the metallic AFM tip. The optical field is localized to a nanoscale region that is smaller than the radius of the probe. The numerical simulation suggests that the infrared field is amplified by the total internal reflection geometry and the tip-enhancement by a factor of $1\times10^4$ for a thin layer of *h*-BN of 70 nm thickness. The localized and enhanced infrared field is capable of exciting PhPs in 2D materials with high spatial frequencies.

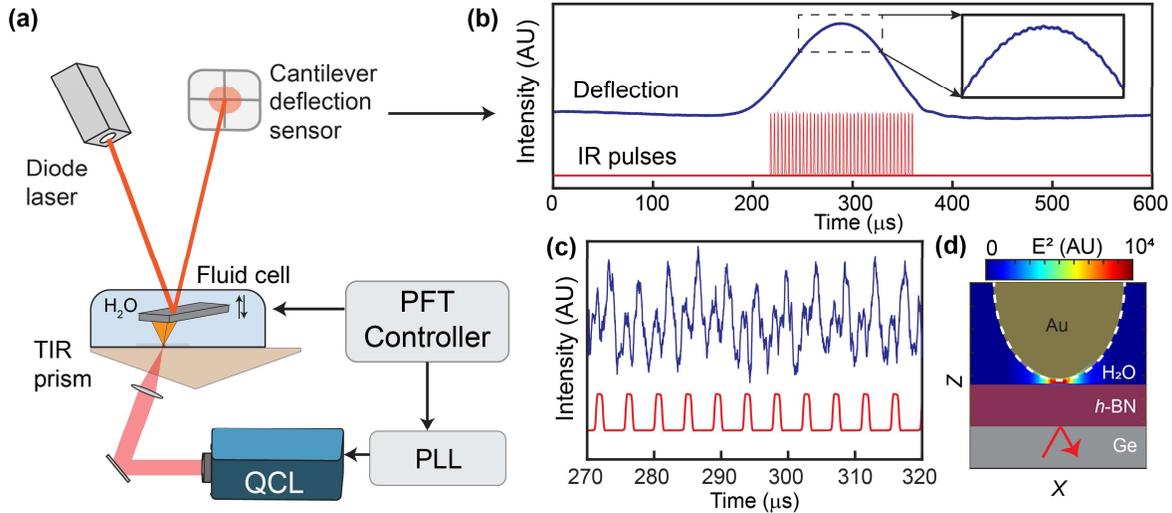

**Figure 1.** The working mechanism of LiPFIR and tip-enhancement under total internal reflection conditions. (a) Schematics of the LiPIFR apparatus. PFT stands for peak force tapping, and PLL stands for phase lock loop. (b) The cantilever deflection traces with photothermal expansions (blue) and a train of laser pulses (red). Inset shows a zoom-in region of cantilever deflection, which contains mechanical perturbation caused by the photothermal expansions. (c) The extracted additional cantilever deflection trace (blue curve) and corresponding laser pulses (red). (d) FDTD simulation of the field confinement of total internal reflection at 20° incident angle (shown by a red arrow) and 1400 cm$^{-1}$. The sample is a 100-nm thick *h*-BN flake, and the Au tip has a radius of 30 nm. The tip-sample distance is set at 1 nm. A field enhancement factor of $10^4$ is achieved between the tip and sample by the tip-enhancement and the evanescent field from total internal reflection.

**Results**

A flake of isotopically enriched hexagonal boron nitride (*h*-$^{10}$BN), which supports hyperbolic phonon polaritons was used for the LiPFIR experiment.[6, 8] The PhPs of *h*-BN are typically probed in the air with the scattering near-field s-SNOM technique. The characteristic feature of polaritons is the interference fringes generated between tip-launched propagating PhPs and their reflection by an edge.[14-15] The fringes reveal the spatial frequencies of the PhPs that can be correlated with the energy of the infrared photons to derive the dispersion relation. Besides s-SNOM, AFM-based photothermal infrared microscopy has also revealed the PhPs in *h*-BN by detecting fringes.[5, 16-18]. However, no prior aqueous phase mid-infrared nano-imaging of PhPs in *h*-BN has been explored in the aqueous phase due to various challenges, from high light delivery loss to anharmonicity in cantilever oscillation in water. Here, we present the first *in situ* nano-imaging of PhPs of *h*-BN in the aqueous phase with LiPFIR.

Figure 2a displays the topography of an edge of an exfoliated $h$-$^{10}$BN flake submerged in distilled water on the surface of a germanium prism. The thickness of the flake is 70 nm. LiPFIR detects characteristic interference fringes in water from 1405 cm$^{-1}$ to 1415 cm$^{-1}$, Figure 2b-d. Above 1420 cm$^{-1}$ the fringe contrast becomes too weak to detect (Figure 2e). The spacing of the interference fringes of PhPs decreases as the frequency of the infrared radiation increases, which is consistent with the general trend of the momentum-energy dispersion relationship of phonon polaritons. The averaged line scans of the interference fringes from 1400 to 1415 cm$^{-1}$ are displayed in Figure 2f, from which a dispersion relation can be extracted.

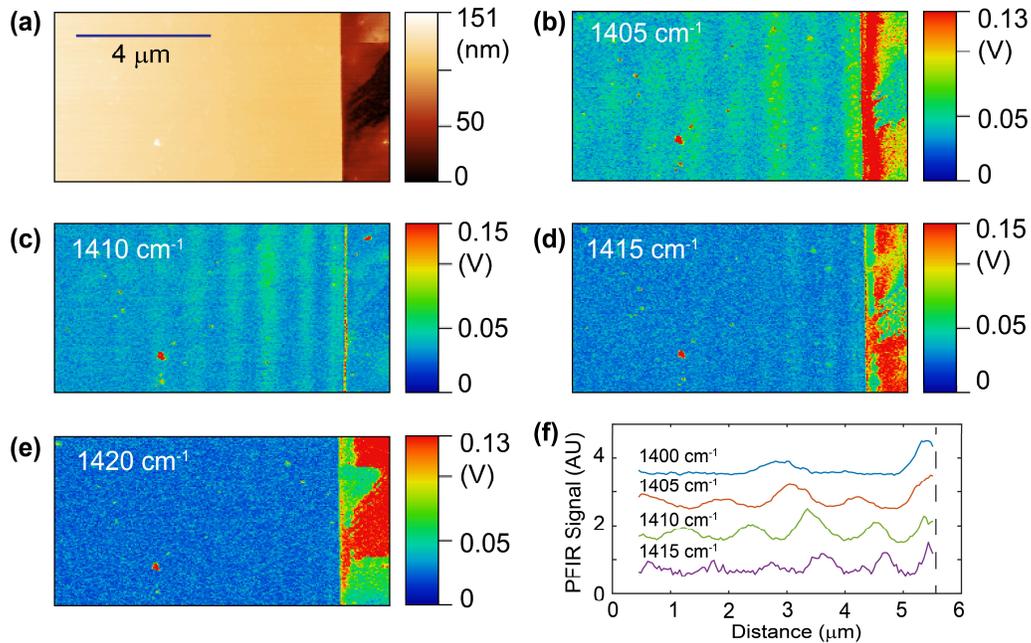

**Figure 2.** Phonon polaritons revealed by LiPFIR. (a) AFM topography of an $h$-$^{10}$BN flake. (b-e) LiPFIR images at 1405, 1410, 1415 and 1420 cm$^{-1}$, respectively. Interference fringes of phonon polaritons are noticeable in (b-d). (f) Averaged line scan profiles of phonon polariton fringes at 1400-1415 cm$^{-1}$, from which the momentum of polariton can be extracted. The position of the edge is shown as the vertical dashed black line.

Infrared spectra are collected with the LiPFIR apparatus by placing the AFM tip at specific locations on the $h$-$^{10}$BN flake. The AFM tip placed on the $h$-$^{10}$BN flake serves as an optical antenna, coupling additional infrared energy into the $h$-$^{10}$BN at that position. The infrared energy—coupled into the polariton-active material by the metallic AFM tip—is eventually turned into heat, causing a mechanically detectable photothermal expansion. Figure 3 displays a series of LiPFIR spectra on the $h$-

$^{10}$BN flake. The infrared absorption spectra transduced by photothermal expansion signals varies with position. The main absorption peak is at 1405 cm$^{-1}$, which is blue-shifted from the $h$-$^{10}$BN phonon resonance at 1395 cm$^{-1}$.[19] The blue shift suggests that the spectroscopic response is dominated by the phonon polaritons rather than the regular phonon resonance, as the former is associated with a higher optical density of states. The change of the LiPFIR spectra with the distance from the edge also suggests the presence of the polaritonic response. As the positions where the LiPFIR spectra are taken progress inward from the edge, the spectral response at 1405 cm$^{-1}$ is maximized at 0.5 $\mu m$, drastically decreases at 1 $\mu m$, then gradual increases towards 2.5 $\mu m$. The signal modulation is due to the spatial distributions of polaritonic fringes. In the absence of the phonon polaritons, the spectra should be constant across a uniform material, rather than exhibiting modulations.

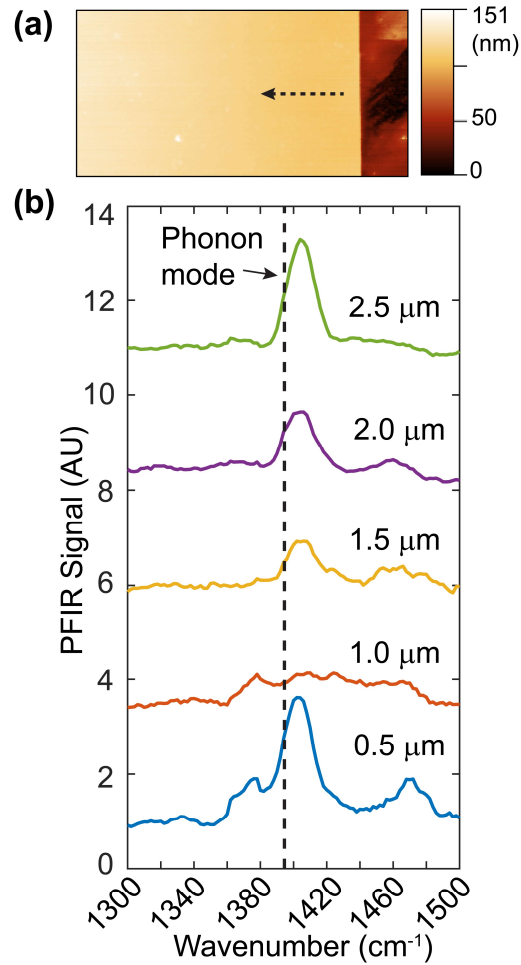

**Figure 3.** Point spectra collected by LiPFIR. (a) AFM topography of an $h$-$^{10}$BN flake. Spectra are collected at five different locations with a 500 nm interval along the black dashed arrow. The first point is 500 nm away from the edge. (b) Point spectra at five different locations on $h$-$^{10}$BN, which are offset vertically for the clarity for comparison. The black dashed line at 1395.4 cm$^{-1}$ shows the intrinsic phonon mode of $h$-$^{10}$BN.[19]

The energy-momentum dispersion relationship of PhPs of $h$-$^{10}$BN is extracted through Fourier transforms of the spatial profiles in Figure 2f into the spatial frequency domains, and then correlate with the energy of the infrared excitation field. Figure 4 displays the experimentally extracted dispersion relation of PhPs in water (blue dots). The theoretical trend of the hyperbolic PhP dispersion relation in water is included as a false-colored map as a reference. The numerical simulation follows the model described in the literature.[6] The photothermally detected PhP in $h$-$^{10}$BN flake in water matches the theoretical model. We have also performed a regular PFIR experiment in the ambient air condition for the same $h$-$^{10}$BN edge. The dispersion relations are extracted and plotted as white dots in Figure 4. There is a clear difference between the PhPs in the air and PhPs in water, suggesting that the dispersion relation of PhPs depends on the surrounding dielectric environment. The origin of this difference forms the basis of chemical sensing with polaritons.

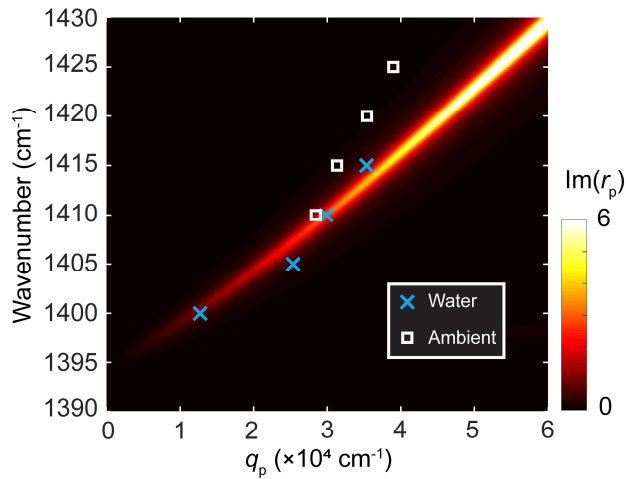

**Figure 4.** Dispersion of hyperbolic polaritons overlaid with LiPFIR data. Experimental data extracted from Fig. 2f are shown as blue crosses. Data collected in ambient conditions are shown as white squares. Note that the PhPs of $h$-$^{10}$BN in the air are not observable below 1410 cm$^{-1}$; whereas PhPs of $h$-$^{10}$BN in water cannot be detected above 1415 cm$^{-1}$. Calculated dispersion relation of $h$-$^{10}$BN phonon polaritons in water is represented by the imaginary part of the reflectivity with a false-color map, assuming a H$_2$O/$h$-$^{10}$BN/Ge structure.[6]

**Discussion**

The imaging capability of LiPFIR is expected to improve with further instrument development. The Bioscope Catalyst AFM that was used in this demonstration has a noise level of r.m.s. 80 pm. The next generation of Bioscope AFM, the Bioscope Resolve, has an improved noise level of r.m.s. 40 pm. The peak force tapping frequency of Bioscope Catalyst is limited to 1 kHz in the fluid, whereas the Bioscope Resolve allows 2 kHz peak force tapping frequency in the liquid. These two factors would straightforwardly increase the signal to noise ratio by an estimated factor of $2\sqrt{2}$ if a Bioscope Resolve AFM will be used instead of the current Bioscope Catalyst AFM.

In our experiment, a simple edge of $h$-$^{10}$BN was measured to demonstrate that PhPs can be detected in the water. However, the simple edge of the $h$-$^{10}$BN does not possess further geometrical-confinement of PhPs, for the lack of conditions to sustain stronger standing waves. The strength of the PhP resonance can be further enhanced with specially tailored shapes, such as a circular geometry resonator,[20-22] an antenna that localizes electrical fields,[23] or a nanofocusing tapered slab.[24] Heterostructures of $h$-BN with other metallic or plasmonic materials[25-26] may also enhance the PhPs in the liquid phase to improve the signal strength. Our work opens a new avenue for those applications and the development of polaritonics in the aqueous phase.

In summary, we demonstrate the feasibility of LiPFIR to image the mid-IR phonon polaritons directly in the aqueous phase. The capability paves the way for *in situ* evaluation of mid-infrared phonon polaritonics in the fluid phase, where the mass transportation capability is favorable for molecular sensing, chemical reactions, and biological transformations.

**Methods and Materials**

**LiPFIR setup**. LiPFIR setup in Fig.1 (a) is composed of an AFM (Bruker Bioscope Catalyst, with Nanoscope V controller and Nanoscope 9.1 software), a quantum cascade laser (MIRcat-QT, DRS Daylight Solutions), and a data acquisition (DAQ) card (PXI-5122, National Instruments). The synchronization between laser pulses train and peak force tapping is realized by a phase lock loop generated from a lock-in amplifier (MFLi, Zurich Instruments). A function generator (HDG2012B, Hantek) working in the burst mode is used to trigger laser pulses. A customized LabView script (LabView 2015, National Instruments) is used to record, process, and output LiPFIR data through another DAQ device (PXI-4461, National Instruments) in real-time for the imaging. Spectral data are further processed by a customized MATLAB script (MATLAB R2019a, MathWorks).

**LiPFIR measurement**. The AFM tip used in this work is a gold-coated tip (HQ:NSC14/Cr-Au, Mikromasch). For LiPFIR measurement in water, 20-40 µL of deionized water (18.2 MΩ) is used to immerse both cantilever and sample. In peak force tapping, a 150-nm amplitude and an 8 nN setpoint are used. The pulse train used in LiPFIR contains 25 pulses with a repetition rate of 530 kHz, to maximize the deflection signals of the AFM cantilever.

***h*-¹⁰BN fabrication and preparation**. The synthesis of *h*-¹⁰BN followed a method reported by Liu *et al*[27] and consisted of two steps: ingot formation and crystal growth. In the ingot formation step, an alumina boat was filled with powdered boron and metal. 2.15 wt% ¹⁰B with balance iron with a total mass of 50 g was used. The alumina boat was put in an alumina tube furnace with a nitrogen purge to remove oxygen, and then a $N_2/H_2$ mixture with 11% $H_2$ is flowed through the tube for the duration of the experiment. The furnace was heated to 1550°C and maintained for 24 hours to ensure the materials melted and mixed well. Afterward, the system is quenched to form an ingot. In the crystal growth step, the ingot was purged with nitrogen, then held the furnace at 1550°C for 24 hours with the same $N_2/H_2$ mixture. Afterward, the furnace was slowly cooled at 1°C/hour to cause *h*-BN to precipitate. Once the furnace reached 1500°C, it is quenched. The resulting ingot was covered in a thin layer of *h*-¹⁰BN crystals and peeled off for usage. The *h*-¹⁰BN thin films on a Ge prism were prepared by exfoliation using a Scotch tape. The Ge prism was then rinsed with acetone before use.

**Simulation**. FDTD simulation in Figure 1(d) was done by Lumerical FDTD (Lumerical Inc.). In the simulation, a Au tip with the end radius of 30 nm was placed above a 100-nm thick *h*-¹⁰BN film, which sits on the Ge prism. A P-polarized (polarized on the same plane as Figure 1(d)) plane wave with a 20° incident angle traveled from the bottom towards the Ge/*h*-BN interface, while the surrounding of the tip-sample region is $H_2O$. The dispersion relation calculated in Figure 4 used model introduced in the literature[6] and was based on a heterostructure of $H_2O$/*h*-¹⁰BN/Ge, details can be found in our previous work.[20] Permittivities of 1.36 (for $H_2O$) and 15.7 (for Ge) and the thickness of 50 nm were used in the simulation. Parameters for *h*-¹⁰BN were from the literature.[19]

**Conflicts of interest**

The authors declare no conflict of interest.

**Acknowledgments**

X. G. X. would like to thank the support from Beckman Young Investigator Award from the Arnold and Mabel Beckman Foundation and the Sloan Research Fellowship from the Alfred P. Sloan Research Foundation. H. W. and X.G.X. would like to thank the support from the National Science Foundation, award number CHE 1847765.

**Author Contributions**

X.G.X. designed the experiment. X.G.X. and H.W. built the LiPFIR experimental apparatus. H. W. carried out the experiment, the collection and analysis of data. E. J. and J. E. provided the $h$-$^{10}$BN sample. H.W. and X.G.X. participated in writing the manuscript. X.G.X. guided the overall research.